\documentstyle[aps,twocolumn,epsf]{revtex}

\begin{document}
\draft

\title{Elastic and Inelastic Evanescent-Wave Mirrors for Cold Atoms}

\author{D. Voigt, B.T. Wolschrijn, R.A. Cornelussen, R. Jansen, N. Bhattacharya,\\
	H.B. van Linden van den Heuvell, and R.J.C. Spreeuw}

\address{Van der Waals-Zeeman Institute, University of Amsterdam, \\
         Valckenierstraat 65, 1018 XE Amsterdam, the Netherlands\\
         http://www.science.uva.nl/research/aplp/}

\date{\today}
\maketitle

\begin{abstract}

We report on experiments on an evanescent-wave mirror for cold $^{87}$Rb
atoms. Measurements of the bouncing fraction show the importance of the
Van der Waals attraction to the surface. We have directly observed
radiation pressure parallel to the surface, exerted on the atoms by the
evanescent-wave mirror. We analyze the radiation pressure by imaging the
motion of the atom cloud after the bounce. The number of photon recoils
ranges from 2 to 31. This is independent of laser power, inversely
proportional to the detuning and proportional to the evanescent-wave
decay length. By operating the mirror on an open transition, we have
also observed atoms that bounce inelastically due to a spontaneous Raman
transition. The observed distributions consist of a dense peak at the
minimum velocity and a long tail of faster atoms, showing that the
transition is a stochastic process with a strong preference to occur
near the turning point of the bounce. 

\end{abstract}


%\narrowtext

\section{Introduction}

The use of evanescent waves (EW) as a tool to manipulate the motion of
neutral atoms has been proposed by Cook and Hill \cite{CooHil82}. Since
then, EW mirrors have become an important tool in atom optics
\cite{AdaSigMly94}. They have been demonstrated for atomic beams at
grazing incidence \cite{BalLetOvc87} and for ultracold atoms at normal
incidence \cite{KasWeiChu90}. In many experiments the scattering of EW
photons was undesirable because it makes the mirror incoherent. 

However, an EW mirror is also a promising tool for efficient loading of
low-dimensional optical atom traps in the vicinity of the dielectric
surface \cite{DesDal96,PowPfaWil97,GauHarSch98,SprVoiWol00}. In these
schemes, spontaneous optical transitions play a crucial role in
providing dissipation \cite{DesArnSzr96,OvcManGri97,LarOvcBal97}. Since
inelastic bouncing may increase the atomic phase-space density, this may
open a route towards quantum degenerate gases, which does not use
evaporative cooling. Thus one may hope to achieve ``atom lasers''
\cite{BECAtomLasers} which are open, driven systems out of thermal
equilibrium, similar to optical lasers \cite{SprPfaJan95}. It is this
application of EW mirrors which drives our interest in experimental
control of the photon scattering of bouncing atoms.

In a first experiment involving the scattering of evanescent photons we
have observed directly for the first time the radiation pressure exerted
by evanescent waves on cold atoms \cite{VoiWolJan00}. In a second
experiment we directly observe clouds of atoms which bounce
inelastically by changing their hyperfine ground state.

\section{Experimental setup}

\begin{figure}[t]
  \centerline{\epsfxsize=7cm\epsffile{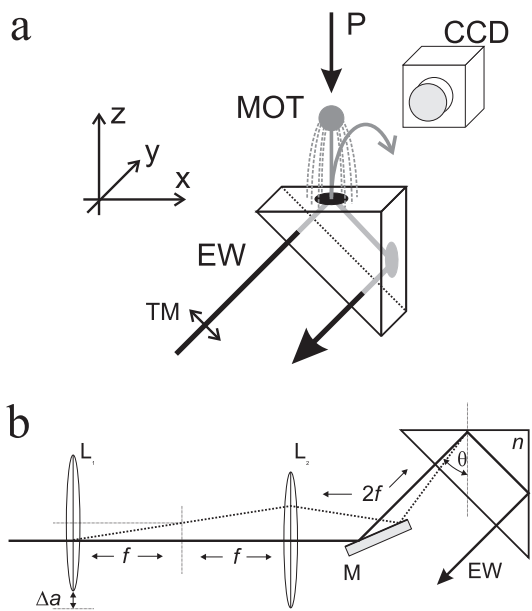}}
  \vspace*{0.5cm}
\caption{Experimental setup. (a) Cold $^{87}$Rb atoms (MOT), are
released 6.6~mm above a right-angle prism. An evanescent wave is created
by beam EW. Fluorescence from probe beam P
is imaged onto a CCD camera. (b) Confocal relay telescope for adjusting
the angle of incidence $\theta$. The lenses $L_{1,2}$ have equal focal
length, $f=75~$mm. A translation of $L_1$ by a distance $\Delta a$
changes the angle of incidence by $\Delta\theta=\Delta a/{f n}$. The
position of the EW spot remains fixed. M is a mirror.}
  \label{fig:SetupFigure}
\end{figure}

Our experiments are performed in a vapor cell. Approximately $10^{7}$
atoms of $^{87}$Rb are loaded out of the background vapor into a
magneto-optical trap (MOT) and are subsequently cooled to {10~$\mu$K}
using polarization gradient cooling (optical molasses). The cold atom
cloud is released in the $F=2$ ground state. After a free fall of 6.6~mm
the atoms reach the horizontal surface of a right-angle BK7 prism
(refractive index $n=1.51$), see Fig.~\ref{fig:SetupFigure}(a).

The EW beam emerges from a single-mode optical fiber, is collimated and
directed into the prism through a relay telescope, see
Fig.~\ref{fig:SetupFigure}(b). The angle of incidence $\theta$ is
controlled by the vertical displacement $\Delta a$ of the first lens
$L_1$. The second lens $L_2$ images the beam to a {\em fixed} spot at
the prism surface. A displacement $\Delta a$ changes the angle by
$\Delta\theta=\Delta a/{n f}$. The beam has a minimum waist of
{$335~\mu$m} at the surface ($1/e^2$ intensity radius) and a nearly
diffraction limited divergence half-angle of $< 1$~mrad.

For the EW, an injection-locked single mode laser diode provides up to
28~mW of optical power behind the fiber. It is seeded by an external
grating stabilized diode laser, locked to the $^{87}$Rb hyperfine
transition $5S_{1/2}(F=2)\rightarrow 5P_{3/2}(F'=3)$ of the $D_2$ line. The
detuning with respect to this transition determines the optical
potential for atoms released from the MOT in
the $F=2$ ground state.

\section{Van der Waals surface attraction}

The EW field is $E\left({\rm\bf r}\right)\propto\exp(ik_x x)\exp(-\kappa
z)$, with $k_x=k_0 n \sin \theta$ and $\kappa=k_0\sqrt{n^2\sin
^{2}\theta -1}$, where $k_0=2\pi/\lambda_0$ is the vacuum wave number.
The optical dipole potential for a two-level
atom can be written as 
${\cal U}_{\rm dip}(z)={\cal U}_0 \exp(-2\kappa z)$. In the limit of large detuning,
$\left|\delta\right|\gg\Gamma$, and low saturation, $s_0\ll 1$, the
maximum potential at the surface is ${\cal U}_0=\hbar\delta s_0$/2, with
a saturation parameter $s_0\simeq(\Gamma/{2\delta})^2\,T I/{I_0}$
\cite{B:CohDupGry92}. Here $I$ is the intensity of the incident beam
inside the prism, $I_0=1.65$~mW/cm$^2$ is the saturation intensity for
rubidium and {$\Gamma=2\pi\times 6.0~$MHz} is the natural linewidth. The
factor $T$ ranges from 5.4 -- 6.0 (2.5 -- 2.65) for a TM (TE)
polarized EW \cite{Fresnel}. The detuning of the laser
frequency $\omega_L$ with respect to the atomic transition frequency
$\omega_0$ is defined as $\delta=\omega_L -\omega_0$. Thus a ``blue''
detuning ($\delta>0$) yields an exponential potential barrier for
incoming atoms. A classical
turning point exists if the barrier height exceeds the
kinetic energy $p_i^2/2M$ of an incident atom.

In reality, the barrier height is determined not only by the dipole
potential. The attractive Van der Waals potential near the surface, 
\begin{equation}
  {\cal U}_{\rm vdW}(z)=
  -\frac{3(n^2-1)}{16(n^2+1)}
  \frac{\hbar\Gamma}{(k_0 z)^3}.
\end{equation}
lowers the maximum potential and thus decreases the
surface area on which atoms can bounce \cite{LanCouLab96}. Gravity can
usually be neglected on the length scale of the EW decay length. 

\begin{figure}[t]
  \centerline{\epsfxsize=8cm\epsffile{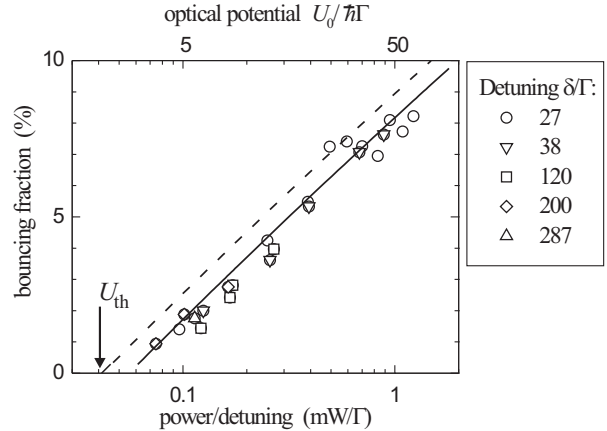}}
  \caption{Bouncing fraction vs. evanescent-wave power and detuning: TE-polarisation,
  power between $0-28$~mW, detuning in units of~$\Gamma=2\pi\times6$~MHz,
  angle $\theta=\theta_{\rm c}+8.7~$mrad, laser waist $w_{\rm 0}=335~\mu$m.
  Predictions with (solid line) and without (dashed) {Van der Waals}
  interaction. The (optical) threshold potential is indicated by an arrow.
  \label{fig:ewfrac}}
\end{figure}

We have measured the fraction of bouncing atoms using a weak resonant
probe beam, inserted between the MOT position and the surface.
Time-of-flight signals were recorded on a photodiode and the integrated
absorption signal was measured for both falling and bouncing atoms. The
extracted fractions of bouncing atoms are shown in
Fig.~\ref{fig:ewfrac}, as a function of the strength of the optical
potential ${\cal U}_0$. The logarithmic dependence is a consequence of
the gaussian beam profile of the EW. Also shown are two calculations of
the bouncing fraction using no adjustable parameters, ({\em i}) assuming
a purely optical potential (dashed line), and ({\em ii}) taking into
account also the Van der Waals surface attraction (solid line). The
latter clearly yields much better agreement. Similar measurements have
previously been performed by Landragin {\em et al.} \cite{LanCouLab96}.

\section{Radiation pressure by evanescent waves}

The evanescent wavevector ${\bf k}=(k_x,0,i\kappa)$ contains 
a real (propagating) component along the surface. 
It has been predicted already by Roosen and Imbert \cite{RooImb76} that 
therefore an evanescent wave should exert radiation pressure parallel to 
the surface. A similar scattering
force has been observed for micrometer-sized
dielectric spheres \cite{KawSug92}. 
The photon scattering rate of a two-level atom
in steady state at low saturation is
$\Gamma'\approx s\Gamma/2=(\Gamma/\hbar\delta)\,{\cal U}_{\rm dip}$
\cite{B:CohDupGry92}. An atom bouncing on an EW mirror sees a time
dependent saturation parameter $s(t)$. Assuming that the excited state
population follows adiabatically, we can integrate the scattering rate
along an atom's trajectory to get the number of scattered photons,
$N_{\rm scat}=\int \Gamma'(t) dt$. For a purely optical potential 
this leads to an analytical solution:
\begin{equation}
  N_{\rm scat}=\frac{\Gamma}{\delta}\frac{p_i}{\hbar\kappa}.
     \label{eq:AnalyticalSolution}
\end{equation}
Note that $N_{\rm scat}$ is independent of ${\cal U}_0$, as a
consequence of the exponential shape of the potential.

We have observed this evanescent-wave radiation pressure directly by
imaging the bouncing atom clouds \cite{VoiWolJan00}. The fluorescence
induced by a 0.5~ms pulse of resonant probe light was imaged onto a
digital frame-transfer CCD camera. A repumping beam, tuned to the
$F=1\rightarrow F'=2$ transition of the $D_1$ line (795~nm), was used to
counteract optical pumping to the $F=1$ ground state by the probe. 

\begin{figure}[t]
  \centerline{\epsfxsize=8.0cm\epsffile{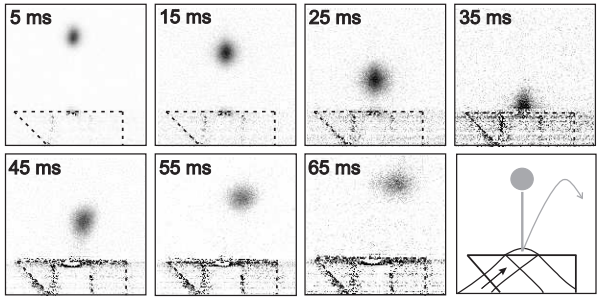}}
  \vspace*{0.5cm}
  \caption{Fluorescence images of a bouncing atom cloud.
	The first image was taken {5~ms} after releasing the atoms from the MOT.
	The configuration of prism and evanescent wave is illustrated by the
	schematic (Field of view: {$10.2\times 10.2~$mm$^2$}).}
  \label{fig:TimeSequence}
\end{figure}

We measure the trajectories of bouncing atoms by taking a sequence of
images with incremental time delays. A typical series with increments of
10~ms is shown in Fig.~\ref{fig:TimeSequence}. A new sample was prepared
for each shot; each image was averaged over 10 shots. In the lower half
of the Figure we see the atom cloud bouncing up slightly sideways. 

\begin{figure}[h]
 \centerline{\epsfxsize=9.0cm\epsffile{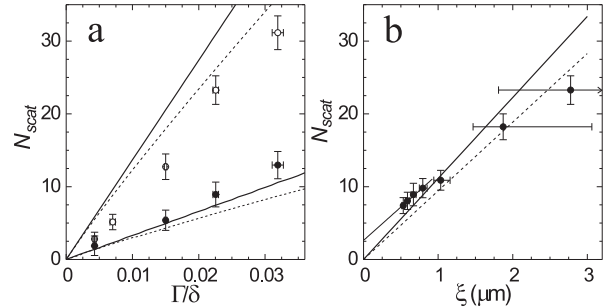}}
 \caption{Radiation pressure on bouncing atoms expressed as number of absorbed
	photons, $N_{\rm scat}$.
    (a)	Detuning $\delta$ varied for {$\xi=2.8~\mu$m} (open points) and
	{$0.67~\mu$m} (solid points).
    (b)	EW decay length $\xi$ varied for $\delta =44~\Gamma$.
	The laser power was {19~mW}. The thin solid line is a linear fit
	through the first four data points.
	Theoretical predictions: two-level atom
	(see Eq.~(\ref{eq:AnalyticalSolution}), thick solid lines).
	Excited-state hyperfine structure and saturation taken into
	account	(dashed lines).}
\label{fig:Recoils}
\end{figure}

The horizontal velocity change during the bounce is extracted from a
sequence of images. In Fig.~\ref{fig:Recoils}, we show how this velocity
kick depends on the laser detuning $\delta$ and on the EW decay length
$\xi\equiv 1/\kappa(\theta)$, by varying the angle $\theta$. The velocity
kick has been expressed in units of the EW photon recoil, $p_{rec}=\hbar
k_0 n\sin\theta$, with $\hbar k_0/M=5.88$~mm/s. In
Fig.\,\ref{fig:Recoils}(a), the detuning is varied from 188 -- 1400~MHz
(31 -- 233~$\Gamma$). Two sets of data are shown, taken for two
different EW decay lengths $\xi=2.8~\mu$m and $0.67~\mu$m.
We find that $N_{\rm scat}\propto\delta^{-1}$, as expected. The
predictions based on Eq.~(\ref{eq:AnalyticalSolution}) are indicated in
the figure (solid lines).

In Fig.~\ref{fig:Recoils}(b), the detuning was kept fixed at $44~\Gamma$
and the angle of incidence varied from 0.9 -- 24.0~mrad above the
critical angle. This leads to a variation of decay length $\xi$ from 2.8
-- 0.53~$\mu$m. Here also, we find a linear dependence on $\xi$. 
Clearly, a steep optical potential, {i.e.} a small decay length,
causes less radiation pressure than a shallow potential. A linear fit to
the data for {$\xi<1~\mu$m}, extrapolates to an offset of approximately
3 photon recoils in the limit $\xi\rightarrow 0$ [thin solid line in
Fig.~\ref{fig:Recoils}(b)]. We attribute this offset to diffusely
scattered EW light due to roughness of the prism surface which
propagates into the vacuum. Preferential light scattering in the
direction of the EW propagation can be explained if the power spectrum
of the surface roughness is narrow compared to $1/\lambda$
\cite{HenMolKaiA97}. The effect of surface roughness has previously been
observed as a broadening of bouncing atom clouds by
the roughness of the {\em dipole potential} \cite{LanLabHen96}. 
In our case, we observe a
change in center of mass motion of the clouds due to an increase in the
{\em spontaneous scattering force}. Such a contribution to the radiation
pressure due to surface roughness vanishes in the limit of large
detuning $\delta$. Thus, we find no significant offset in
Fig.~\ref{fig:Recoils}(a). 

We have also verified that there is no significant dependence on the
power of the EW, in accordance with
Eq.~(\ref{eq:AnalyticalSolution}). Only a slight increase with EW power
is observed, which may again be due to diffusely scattered light. The
optical power mainly determines the effective mirror surface and thus
the fraction of bouncing atoms. This is also visible in the horizontal
width of bouncing clouds.

We find a significant correction due to the excited state manifold
$F'=\{0,1,2,3\}$ of {$^{87}$Rb}. Besides $F'=3$, also $F'=1,2$ contribute
to the mirror potential, whereas they do not much affect the scattering
rate. For an EW detuning of $44\,\Gamma$, this reduces the number of
scattered photons by typically $9\,\%$ compared to a two-level atom,
when we average over the contributions from different magnetic
sublevels (dashed lines).

\section{Inelastic bouncing}

If the mirror laser is tuned to an open transition, photon scattering
can lead to a spontaneous Raman transition to the other hyperfine ground
state. The atom will then leave the surface on a different potential
curve from when it approached the surface. Hence, the kinetic energy
after the bounce will in general be different from that before the
bounce. 

We performed experiments where the EW laser was tuned in the blue wing
of the $F=1\rightarrow F'=2$ transition of the $D_2$ line of $^{87}$Rb,
with a detuning $\delta_1\ll\delta_{\rm GHF}$, with $\delta_{\rm
GHF}=2\pi\times 6.8$~GHz the ground state hyperfine splitting. Atoms in
$F=2$ see the EW mirror with a detuning which is larger by a factor of
approximately $\delta_2/\delta_1\approx\delta_{\rm GHF}/\delta_1$ and
therefore see a much weaker potential. Thus they leave the surface with
a reduced kinetic energy. Repeating this procedure leads to so-called
``evanescent-wave cooling'' \cite{DesArnSzr96,OvcManGri97,LarOvcBal97}. 

\begin{figure}[h]
 \centerline{\epsfxsize=4cm\epsffile{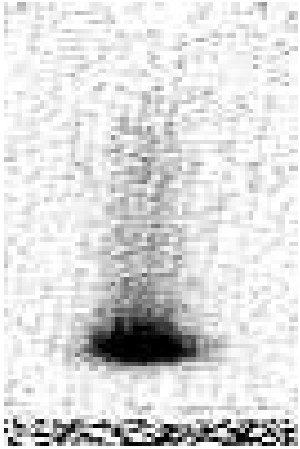}}
 \vspace*{0.5cm}
\caption{Absorption image of atoms bouncing inelastically from an
evanescent-wave mirror. The image was taken 14~ms after the bounce. The
field of view is  $3.4\times 5.1$~mm$^2$.}
\label{fig:inelastic}
\end{figure}

We observe the inelastically bouncing atoms directly by imaging the
absorption of a weak resonant probe pulse onto a digital CCD camera. In
Fig.~\ref{fig:inelastic} we show a typical image, taken when the slowest
atoms have reached their upper turning point. The observed clouds are
highly asymmetric, with a dense peak at the lower edge of the cloud and
a tail of faster atoms extending upward. The dense peak contains the
atoms that made the Raman transition near the turning point, because
here they move slowly in a relatively high light intensity. The tail
contains faster atoms, which have made the Raman transition while
falling down to, or bouncing up from the surface.
The density distribution in the vertical direction agrees well with a
calculation where the atoms are modeled as point particles moving on an
exponential potential curve.

\section{Outlook}

The result of the inelastic bouncing experiment shows the preference for
making a spontaneous Raman transition near the turning point of the
bounce. This is of great importance for our ongoing experiments which
are aimed at trapping the cold atoms near the turning point, in a
low-dimensional optical trap. Although the present experiment showed a
relatively high density at the lowest possible bouncing velocity, it is
unlikely that the phase-space density has been increased. It remains to be
investigated under what conditions such an increase can be obtained and
by how much. 

The scattering of evanescent photons is an essential ingredient in our
experiment toward loading low-dimensional optical traps. At the same
time, once loaded into the trap, we wish the atoms to scatter at a very
low rate. In order to meet these conflicting requirements our present
experiments are aimed at trapping the atoms in a so-called dark state
which no longer interacts with the evanescent wave. This requires the
use of circularly polarized evanescent waves. This procedure should lead
to an ultracold low-dimensional gas and may ultimately yield an
all-optical road to quantum-degeneracy and an atom laser.

\section*{Acknowledgments}

We thank E.C. Schilder for help with the experiments. This work is part
of the research program of the ``Stichting voor Fundamenteel Onderzoek
der Materie (FOM)'' which is financially supported by the ``Nederlandse
Organisatie voor Wetenschappelijk Onderzoek (NWO)''. R.S. has been
financially supported by the Royal Netherlands Academy of Arts and
Sciences.

\bibliographystyle{prsty}

%\bibliography{d:/research/bib/amo,d:/research/bib/books}

\end{document}